\documentclass{article} 
\usepackage{color}
\usepackage{amsmath,amsfonts,amsthm,amssymb}
\usepackage[colorlinks=true,urlcolor=webblue,linkcolor=webgreen,filecolor=webblue,citecolor=webgreen,pdfpagemode=UseOutlines,pdfstartview=FitH,pdfpagelayout=OneColumn,bookmarks=true]{hyperref}
\usepackage{fullpage}

\hypersetup{
  pdftitle=The symmetric group defies strong Fourier sampling: Part II,
  pdfauthor=Cristopher Moore and Alexander Russell}

\definecolor{webgreen}{rgb}{0,.5,0}
\definecolor{webblue}{rgb}{0,0,.5}

\DeclareMathOperator{\Exp}{Exp}
\DeclareMathOperator{\Var}{Var}

\numberwithin{equation}{section}

\newtheorem{theorem}{Theorem}
\newtheorem{lemma}[theorem]{Lemma}
\newtheorem{conjecture}{Conjecture}

\newtheorem{corollary}[theorem]{Corollary}
\newtheorem{definition}{Definition}
\DeclareMathAlphabet{\varmathbb}{U}{bbold}{m}{n}
\newcommand{\one}{\varmathbb 1}
\renewcommand{\vec}[1]{\mathbf{#1}}

\newcommand{\remove}[1]{}

\newcommand{\C}{\mathbb{C}}
\newcommand{\R}{\mathbb{R}}
\newcommand{\Z}{\mathbb{Z}}

\newcommand{\ket}[1]{\left| #1 \right\rangle}
\newcommand{\bra}[1]{\left\langle #1 \right|}

\newcommand{\CG}{\C[G]}
\newcommand{\cg}{\frac{1}{\sqrt{|G|}}}

\newcommand{\rank}{\textbf{rk}\;}
\newcommand{\norm}[1]{\left\| #1 \right\|}
\newcommand{\abs}[1]{\left| #1 \right|}
\newcommand{\U}{\textsf{U}}

\newcommand{\inner}[2]{\left\langle #1, #2 \right\rangle}
\newcommand{\supp}{{\rm supp}}
\newcommand{\poly}{{\rm poly}}
\newcommand{\chrangle}[1]{\rangle_{#1}}
\newcommand{\chlangle}{\langle}

\newcommand{\vb}{\vec{b}}

\newcommand{\vu}{\vec{u}}
\newcommand{\vx}{\vec{x}}
\newcommand{\vv}{\vec{v}}
\newcommand{\bad}{B_{\textrm{bad}}}
\newcommand{\reg}{\textrm{R}}
\newcommand{\conj}{\textrm{C}}
\newcommand{\vrho}{{\mathbf \rho}}
\newcommand{\vsigma}{{\mathbf \sigma}}
\newcommand{\oP}{\overline{P}}

\title{The Symmetric Group Defies Strong Fourier Sampling: Part II}

\author{Cristopher Moore \\
  \textsf{moore@cs.unm.edu}\\
  Department of Computer Science\\
  University of New Mexico
  \and
  Alexander Russell\\
  \textsf{acr@cse.uconn.edu}\\
  Department of Computer Science and Engineering\\
  University of Connecticut}

\begin{document}
\maketitle 

\begin{abstract}
  Part I of this paper showed that the hidden subgroup problem over
  the symmetric group---including the special case relevant to Graph
  Isomorphism---cannot be efficiently solved by strong Fourier
  sampling, even if one may perform an arbitrary POVM on the coset
  state.  In this paper, we extend these results to entangled
  measurements.  Specifically, we show that the case of the hidden subgroup
  problem which is relevant to Graph Isomorphism cannot be solved
  by any polynomial number of experiments on one- or two-register coset states.
\end{abstract}

\section{Introduction: the hidden subgroup problem}

Many problems of interest in quantum computing can be
reduced to an instance of the \emph{Hidden Subgroup Problem}
(HSP). This is the problem of determining a subgoup $H$ of a group $G$
given oracle access to a function $f: G \to S$ with the property that
$$
f(g) = f(hg) \Leftrightarrow h \in H\enspace.
$$
Equivalently, $f$ is constant on the cosets of $H$ and takes
distinct values on distinct cosets.

All known efficient solutions to the problem rely on the
\emph{standard method} or the method of \emph{Fourier sampling}
\cite{BernsteinV93}, described below.

\begin{description}
\item[Step 1.] Prepare two registers, the first in a uniform
  superposition over the elements of $G$ and the second with the value
  zero, yielding the state
  $$
  \psi_1 = \cg \sum_{g \in G} \ket{g} \otimes \ket{0} \enspace.
$$
\item[Step 2.] Query (or calculate) the function $f$ defined on $G$
  and XOR it with the second register.  This entangles the two
  registers and results in the state
  $$
  \psi_2 = \cg \sum_{g \in G} \ket{g} \otimes \ket{f(g)} \enspace.
$$
\item[Step 3.] Measure the second register.  This puts the first
  register in a uniform superposition over one of $f$'s level sets,
  i.e., one of the cosets of $H$, and disentangles it from the second
  register.  If we observe the value $f(c)$, we have the state $\psi_3 \otimes
  \ket{f(c)}$ where
$$
\psi_3 = \ket{cH}
= \frac{1}{\sqrt{|H|}} \; \sum_{h \in H} \ket{ch}
 \enspace.
 $$
  
\item[Step 4.] Carry out the quantum Fourier transform on $\psi_3$ and
  measure the result.
\end{description}

The result of Step 3 above is the \emph{coset state} $\ket{cH}$, where
$c$ is chosen uniformly from $G$. Expressing this as a mixed state,
let
\[
\rho_H = \frac{1}{|G|} \sum_{c \in G} \ket{cH}\bra{cH}\enspace.
\]
We shall focus on the \emph{hidden conjugate problem}, where the
hidden subgroup is a random conjugate $H^g = g^{-1}Hg$ of a known
(non-normal) subgroup $H$.  It was shown in the first part of this
paper that when $G = S_{2n}$, the symmetric group on $2n$ letters, and
$H$ is a the subgroup generated by the involution
$(1\;2)\ldots(2n-1\;2n)$, the outcome of \emph{any} measurement on $\rho_H$
is nearly independent of the random choice of $g \in S_n$. In
particular, no polynomial number of coset state experiments can
determine such a hidden subgroup with non-negligible probability.

It is known, however, that a measurement exists to determine hidden
subgroups of a group $G$ from $k = \poly \log |G|$ independent copies
of $\rho_H$. In light of the discussion above, this measurement cannot,
in general, be a product measurement: it must involve \emph{entangled}
measurement operators. In this paper, we extend the framework of part
I to such entangled measurements, showing that for the subgroup $H$ of
$S_{2n}$ described above, the result of any measurement of the
two-coset state $\rho_{H^g} \otimes \rho_{H^g}$ is nearly independent of $g$.
In particular, no polynomial number of two-register coset-state
experiments can determine $H$ with non-negligible probability.

\paragraph{Related work.}
Both Simon's and Shor's seminal algorithms rely on the standard method
over an Abelian group. In Simon's problem~\cite{Simon97}, $G=\Z_2^n$
and $f$ is an oracle such that, for some $y$, $f(x) = f(x+y)$ for all
$x$; in this case $H=\{0,y\}$ and we wish to identify $y$.  In Shor's
factoring algorithm~\cite{Shor97} $G$ is the group $\Z_n^*$ where $n$
is the number we wish to factor, $f(x) = r^x \bmod n$ for a random $r
< n$, and $H$ is the subgroup of $\Z_n^*$ whose index is the
multiplicative order of $r$.  (Note that in Shor's algorithm, since
$|\Z_n^*|$ is unknown, the Fourier transform is performed over $\Z_q$
for some $q=\poly(n)$; see \cite{Shor97} or \cite{HalesH99,HalesH00}.)

For such abelian instances; it is not hard to see that a polynomial
number (i.e., polynomial in $\log |G|$) of experiments of this type
determine $H$.  In essence, each experiment yields a random element of
the dual space $H^\perp$ perpendicular to $H$'s characteristic function,
and as soon as these elements span $H^\perp$ they, in particular,
determine $H$.

While the \emph{nonabelian} hidden subgroup problem appears to be much
more difficult, it has very attractive applications.  In particular,
solving the HSP for the symmetric group $S_n$ would provide an
efficient quantum algorithm for the Graph Automorphism and Graph
Isomorphism problems (see e.g.\ Jozsa~\cite{Jozsa00} for a review).
Another important motivation is the relationship between the HSP over
the dihedral group with hidden shift problems~\cite{vanDamHI03} and
cryptographically important cases of the Shortest Lattice Vector
problem~\cite{Regev}.

So far, algorithms for the HSP are only known for a few families of
nonabelian groups, including wreath products $\Z_2^k\; \wr\;
\Z_2$~\cite{RoettelerB98}; more generally, semidirect products $K
\ltimes \Z_2^k$ where $K$ is of polynomial size, and groups whose
commutator subgroup is of polynomial size~\cite{IvanyosMS01};
``smoothly solvable'' groups~\cite{FriedlIMSS02}; and some semidirect
products of cyclic groups~\cite{InuiLeGall}.  Ettinger and
H{\o}yer~\cite{EttingerH98} provided another type of result, by showing
that Fourier sampling can solve the HSP for the dihedral groups $D_n$
in an \emph{information-theoretic} sense.  That is, a polynomial
number of experiments gives enough information to reconstruct the
subgroup, though it is unfortunately unknown how to determine $H$ from
this information in polynomial time.

To discuss Fourier sampling for a nonabelian group $G$, one needs to
develop the Fourier transform over $G$. For abelian groups, the
Fourier basis functions are homomorphisms $\phi:G \to \C$ such as the
familiar exponential function $\phi_k(x) = e^{2\pi i kx/n}$ for the cyclic
group $\Z_n$.  In the nonabelian case, there are not enough such
homomorphisms to span the space of all $\C$-valued functions on $G$;
to complete the picture, one introduces \emph{representations} of the
group, namely homomorphisms $\rho:G \to \U(V)$ where $\U(V)$ is the group
of unitary matrices acting on some $\C$-vector space $V$ of dimension
$d_\rho$. It suffices to consider \emph{irreducible} representations,
namely those for which no nontrivial subspace of $V$ is fixed by the
various operators $\rho(g)$.
Once a basis for each irreducible $\rho$ is chosen, the matrix elements
$\rho_{ij}$ provide an orthogonal basis for the vector space of all
$\C$-valued functions on $G$.

\begin{sloppypar}
  The quantum Fourier transform then consists of transforming
  (unit-length) vectors in $\CG = \{ \sum_{g \in G} \alpha_g \ket{g} \mid \alpha_g
  \in \C\}$ from the basis $\{ \ket{g} \mid g \in G \}$ to the basis $\{
  \ket{\rho,i,j} \}$ where $\rho$ is the name of an irreducible
  representation and $1 \leq i, j \leq d_\rho$ index a row and column (in a
  chosen basis for $V$).  Indeed, this transformation can be carried
  out efficiently for a wide variety of
  groups~\cite{Beals97,Hoyer97,MooreRR04}. Note, however, that a
  nonabelian group $G$ does not distinguish any specific basis for its
  irreducible representations which necessitates a rather dramatic
  choice on the part of the transform designer. Indeed, careful basis
  selection appears to be critical for obtaining efficient Fourier
  transforms for the groups mentioned above.
\end{sloppypar}

Perhaps the most fundamental question concerning the hidden subgroup
problem is whether there is always a basis for the representations of
$G$ such that measuring in this basis (in Step 4, above) provides
enough information to determine the subgroup $H$. This framework is
known as \emph{strong Fourier sampling}.  Part I of this article
answers this question in the negative, showing that natural subgroups
of $S_n$ cannot be determined by this process; in fact, it shows this
for an even more general model, where we perform an arbitrary positive
operator-valued measurement (POVM) on coset states $\ket{cH}$.  We
emphasize that this result includes the most important special cases
of the nonabelian HSP, as they are those to which Graph Isomorphism naturally
reduces.  Namely, as in~\cite{HallgrenRT00} we focus on order-2
subgroups of the form $\{1,m\}$, where $m$ is an involution consisting
of $n/2$ disjoint transpositions; then if we fix two rigid connected
graphs of size $n/2$ and consider permutations of their disjoint
union, then the hidden subgroup is of this form if the graphs are
isomorphic and trivial if they are not.

The next logical step is to consider \emph{multi-register} algorithms,
in which we prepare multiple coset states and subject them to
\emph{entangled} measurements.  Ettinger, H{\o}yer and
Knill~\cite{EttingerHK04} showed that the HSP on arbitrary groups can
be solved information-theoretically with a polynomial number of
registers, although their algorithm takes exponential time for most
groups of interest.  Kuperberg~\cite{Kuperberg03} devised a
subexponential ($2^{O(\sqrt{\log n})}$) algorithm for the HSP on the
dihedral group $D_n$ that works by performing entangled measurements
on two registers at a time, and Bacon, Childs, and van
Dam~\cite{BaconCvD} have determined the optimal multiregister
measurement for the dihedral group.

Whether a similar approach can be taken for the symmetric group is a
major open question.  In this paper, we take a step towards answering
this question by showing that if we perform arbitrary entangled
measurements over {\em pairs} of registers, distinguishing $H=\{1,m\}$
from the trivial group in $S_n$ requires a superpolynomial number
(specifically, $e^{\Omega(\sqrt{n}/\log n)}$) of experiments.

\section{Two combinatorial representations}
\label{sec:representation-theory}

With apologies to the reader, we will rely on the introductory sections of 
Part I rather than repeating them here.  However, here we introduce 
two combinatorial representations that will be extremely useful to us.

For a group $G$, we let $\widehat{G}$ denote a collection of unitary
representations of $G$ consisting of exactly one from each isomorphism
class. We let $\CG$ denote the \emph{group algebra} of $G$; this is
the $|G|$-dimensional vector space of formal sums $$
\Bigl\{ \sum_g \alpha_g \cdot g
\mid \alpha_g \in \C \Bigr\} $$
equipped with the unique inner product for which
$\langle g, h\rangle$ is equal to one when $g = h$ and zero otherwise.  (Note that
$\CG$ is precisely the Hilbert space of a single register containing a
superposition of group elements.)

We introduce two combinatorial representations related to the group
algebra.  The first is the \emph{regular} representation $\reg$, given
by the permutation action of $G$ on itself.  Then $\reg$ is the
representation $\reg: G \to \U(\CG)$ given by linearly extending left
multiplication, $\reg(g): h \mapsto gh$.  It is not hard to see that its
character $\chi_\reg$ is given by $$
\chi_\reg(g) = \begin{cases} |G| & g = 1\enspace,\\
  0 & g \neq 1 \enspace ,
\end{cases}
$$
in which case we have $\chlangle \chi_\reg, \chi_\rho \chrangle{G} = d_\sigma$
for each $\rho \in \widehat{G}$.  Thus $\R$ contains $d_\rho$ copies of 
each irreducible $\rho \in \widehat{G}$, and counting dimensions on each side of this 
decomposition implies $|G| = \sum_{\rho \in \widehat{G}} d_\rho^2$.  

The other combinatorial representation we will rely on is the \emph{conjugation}
representation $\conj$, given by the conjugation action of $G$ on
$\CG$. Specifically, $\conj: G \to \U(\CG)$ is the map obtained by
linearly extending the rule $\conj(g): h \mapsto ghg^{-1}$.
While the decomposition of $C$ into irreducibles is, in general,
unknown, one does have the decomposition
\begin{equation}
\conj = \bigoplus_{\rho \in \widehat{G}} \rho \otimes \rho^* 
 \quad \mbox{and therefore} \quad
\chi_\conj(g) = \sum_{\rho \in \widehat{G}} \chi_\rho(g) \chi_\rho(g)^* 
\enspace . 
\end{equation}
Here $\rho^*$ denotes the complex conjugate representation of $\rho$, which
acts on vectors $\vu^*$ as $\rho^*(g)\vu = (\rho(g) \vu)^*$.  We also note that
an elementary argument shows that
$$
\chi_\conj(g) = \frac{|G|}{|[g]|}\enspace,
$$
where $[g] = \{ h^{-1}gh \mid h \in G \}$ denotes the conjugacy class of $g$.

\section{Background from Part I}
\label{sec:background}

\subsection{The structure of the optimal measurement}

As in Part I, we focus on the special case of the hidden subgroup
problem called the \emph{hidden conjugate problem}
in~\cite{MooreRRS04}.  Here there is a (non-normal) subgroup $H$, and
we are promised that the hidden subgroup is one of its conjugates,
$H^g = g^{-1} H g$ for some $g \in G$; the goal is to determine which.

The most general possible measurement in quantum mechanics is a
positive operator-valued measurement (POVM).  Part I of this paper
establishes that the optimal POVM for the Hidden Subgroup Problem on a
single coset state consists of measuring the name $\rho$ of the
irreducible representation, followed by a POVM on the vector space $V$
on which $\rho$ acts.  In the special case of a von Neumann measurement,
this corresponds to measuring the row of $\rho$ in some orthonormal
basis; in general it consists of measuring according to some
over-complete basis, or \emph{frame}, $B=\{ \vb \}$ with positive real
weights $a_\vb$ that obeys the completeness condition
\begin{equation}
\label{eq:complete2}
\sum_\vb a_\vb \pi_{\vb} = \one \enspace ,
\end{equation}
where $\pi_\vb$ denotes the projection onto the unit length vector
$\vb$. We remark that the frame $B$ weighted according to $a$ is
energy-conserving in the sense that
$$
\| \vx \|^2 = \langle \vx, \one \vx \rangle = \langle \vx, \sum_{\vb} a(\vb)
\pi_{\vb}(\vx)\rangle = \sum_{\vb} a_{\vb} \| \pi_{\vb}(\vx)\|^2\enspace.
$$

During Fourier sampling, the probability we observe $\rho$, and the
conditional probability that we observe a given $\vb \in B$, are given
by
\begin{eqnarray}
 P(\rho) & = & \frac{d_\rho |H|}{|G|} \,\rank \Pi_H 
 \label{eq:prho2} \\
 P(\rho,\vb) & = & a_j \frac{\norm{\Pi_H \vb}^2}{\rank \Pi_H} 
\label{eq:prhoi2}
\end{eqnarray}
where $\Pi_H$ is the projection operator $1/|H| \sum_{h \in H} \rho(h)$.
In the case where $H$ is the trivial subgroup, $\Pi_H = \one_{d_\rho}$ and $P(\rho,\vb_j)$ is given by
\begin{equation}
\label{eq:natural}
 P(\rho,\vb) = \frac{a_\vb}{d_\rho} \enspace . 
\end{equation}
We call this the \emph{natural distribution} on the frame $B=\{ \vb \}$.  In the case that $B$ is an orthonormal basis, $a_\vb = 1$ and this is simply the uniform distribution.
This probability distribution over $B$ changes for a conjugate $H^g$ in the following way:
\[ P(\rho,\vb) = a_j \frac{\norm{\Pi_H g \vb}^2}{\rank \Pi_H} \]
where we write $g \vb$ for $\rho(g) \vb$. It is not hard to show that, 
for any $\vb \in V$, the \emph{expected} value of $\norm{\Pi_H(g \vb)}^2$, over the choice of
$g \in G$, is $\rank \Pi_H / d_\rho$.

\subsection{The expectation and variance of an involution projector}

The following lemmas are proved in Part I; we repeat them here for convenience.

\begin{lemma}
\label{lem:exp}
Let $\rho$ be a representation of a group $G$ acting on a space $V$ and
let $\vb \in V$.  Let $m$ be an element chosen uniformly from a
conjugacy class $I$ of involutions.  If $\rho$ is irreducible, then
\[ \Exp_m \langle \vb, m \vb \rangle = \frac{\chi_\rho(I)}{\dim \rho} \norm{\vb}^2 \enspace . \]
If $\rho$ is reducible, then 
\[ \Exp_m \langle \vb, m \vb \rangle 
= \sum_{\sigma \prec \rho} \frac{\chi_\sigma(I)}{\dim \sigma} \norm{\Pi^\rho_\sigma \vb}^2 \enspace . \]
\end{lemma}

\begin{lemma}
\label{lem:second}
Let $\rho$ be a representation of a group $G$ acting on a space $V$ and
let $\vb \in V$.  Let $m$ be an element chosen uniformly at random from a
conjugacy class $I$ of involutions.  Then
\[ \Exp_m \abs{\langle \vb, m \vb \rangle}^2 
= \sum_{\sigma \prec \rho \otimes \rho^*} 
\frac{\chi_\sigma(I)}{\dim \sigma} \norm{\Pi^{\rho \otimes \rho^*}_\sigma (\vb \otimes \vb^*)}^2 
\enspace . \]
\end{lemma}

Given an involution $m$ and the hidden subgroup $H=\{1,m\}$, 
let $\Pi_m = \Pi_H$ denote the projection operator given by
$$
\Pi_m \vv = \frac{\vv + m \vv}{2}\enspace.
$$
Then the expectation and variance of $\norm{\Pi_m \vb}^2$ are given by the following lemma.

\begin{lemma}  
\label{lem:var}
Let $\rho$ be an irreducible representation acting on a space $V$ and
let $\vb \in V$.  Let $m$ be an element chosen uniformly at random from a
conjugacy class $I$ of involutions.  Then
\begin{eqnarray}
  \Exp_{m} \norm{\Pi_m \vb}^2
  & = & \frac{1}{2} \norm{\vb}^2 \left(1 + \frac{\chi_\rho(I)}{\dim \rho} \right) 
  \label{eq:exp} \\
  \Var_{m} \norm{\Pi_m \vb}^2 
  & \leq & \frac{1}{4} 
 \sum_{\sigma \prec \rho \otimes \rho^*} \frac{\chi_\sigma(I)}{\dim \sigma}
   \norm{\Pi^{\rho \otimes \rho^*}_\sigma(\vb \otimes \vb^*)}^2 
  \enspace.
  \label{eq:var} 
\end{eqnarray}
\end{lemma}

Finally, we point out that since
\[ 
\Exp_{m} \norm{\Pi_m \vb}^2 = \norm{\vb}^2 \frac{\rank \Pi_m}{\dim \rho}
\]
we have
\begin{equation}
\label{eq:rank}
\frac{\rank \Pi_m}{\dim \rho} = \frac{1}{2}  \left(1 + \frac{\chi_\rho(I)}{\dim \rho} \right)
\enspace .
\end{equation}

\subsection{The representation theory of the symmetric group}

We will use several specific properties of the symmetric group $S_n$
and its asymptotic representation theory; we refer the reader to
Section~5 of Part I for more background and notation.  Recall that the
irreducible representations $S^\lambda$ of $S_n$ are labeled by Young
diagrams $\lambda$, and that the number of irreducible representations is
the partition number $p(n)$, which obeys
\begin{equation}
\label{eq:pn}
 p(n) < e^{\delta \sqrt{n}} \mbox{ where } \delta = \pi \sqrt{2/3} \enspace . 
 \end{equation}
 We denote the dimension and character of $S^\lambda$ as $d^\lambda$ and $\chi^\lambda$ 
 respectively.  Recall also that the {\em Plancherel distribution} assigns the probability 
 $d_\rho^2 / |G|$ to each irreducible representation $\rho$.  Then we will rely on the
following results of Vershik and Kerov.

\begin{theorem}[\cite{VershikK}]
  \label{thm:plancherel}
  Let $S^\lambda$ be chosen from $\widehat{S_n}$ according to the Plancherel
  distribution. Then there exist positive constants $c_1$ and $c_2$
  for which
  $$
  \lim_{n \to \infty} \Pr\left[ e^{-c_1 \sqrt{n}} \sqrt{n!} \leq d^\lambda \leq
    e^{-c_2 \sqrt{n}} \sqrt{n!} \right] = 1\enspace.
  $$
\end{theorem}

\begin{theorem}[\cite{VershikK}]\label{thm:max-dimension}
  There exist positive constants $\check{c}$ and $\hat{c}$ such that for all $n \geq 1$,
  $$
  e^{-\check{c} \sqrt{n}} \sqrt{n!} \leq 
  \max_{S^\lambda \in \widehat{S_n}} d^\lambda 
  \leq e^{-\hat{c}\sqrt{n}} \sqrt{n!} \enspace .
  $$
\end{theorem}

In Part I we prove the following:

\begin{lemma}
  \label{lem:lower}
  Let $S^\lambda$ be chosen according to the Plancherel distribution on
  $\widehat{S_n}$.
  \begin{enumerate}
  \item\label{item:lower-sqrt} Let $\delta = \pi \sqrt{2/3}$ as in~\eqref{eq:pn}. Then for
    sufficiently large $n$, $\Pr\left[ d^\lambda \leq e^{-\delta \sqrt{n}}
      \sqrt{n!} \right] < e^{-\delta\sqrt{n}}$.
  \item\label{item:lower-linear} Let $0 < c < 1/2$. Then $\Pr[ d^\lambda \leq
    n^{cn} ] = n^{-\Omega(n)}$.
  \end{enumerate}
\end{lemma}

Finally, we will also apply Roichman's~\cite{Roichman96} estimates for
the characters of the symmetric group:
\begin{definition}
  For a permutation $\pi \in S_n$, define the \emph{support} of $\pi$,
  denoted $\supp(\pi)$, to be the cardinality of the set $\{ k \in [n] \mid
  \pi(k) \neq k \}$.
\end{definition}

\begin{theorem}[\cite{Roichman96}] \label{thm:roichman}
  There exist constants $b > 0$ and $0 < q < 1$ so that for $n > 4$,
  for every conjugacy class $C$ of $S_n$, and every irreducible
  representation $S^\lambda$ of $S_n$,
  $$
  \abs{\frac{\chi^\lambda(C)}{d^\lambda}} 
  \leq \left(\max\Bigl(q, \frac{\lambda_1}{n}, \frac{\lambda'_1}{n}\Bigr)\right)^{b \cdot \supp(C)}
  \enspace,
  $$
  where $\supp(C) = \supp(\pi)$ for any $\pi \in C$.
\end{theorem}

In our application, we take $n$ to be even and consider involutions
$m$ in the conjugacy class of elements consisting of $n/2$ disjoint
transpositions, $M = M_n = \{ \sigma \,((12)(34)\cdots(n-1 \enspace n))\, \sigma^{-1} \mid
\sigma \in S_n\}$.  Note that each $m \in M_n$ is associated with one of the
$(n-1)!!$ perfect matchings of $n$ things, and that $\supp(m) = n$.

\subsection{Strong Fourier sampling on one register}

The main result of Part I is the following.

\begin{theorem}
\label{thm:one-frames}
Let $B=\{\vb\}$ be a frame with 
weights $\{a_\vb\}$ satisfying the completeness condition~\eqref{eq:complete2} for an
irreducible representation $S^\lambda$.  
Given the hidden subgroup $H=\{1,m\}$ where $m$ is chosen uniformly
at random from $M$, let $P_m(\vb)$ be the probability
that we observe the vector $\vb$ conditioned on having observed
the representation name $S^\lambda$, and let $N$ be the natural 
distribution~\eqref{eq:natural} on $B$.  Then there is a constant $\delta > 0$ such that 
for sufficiently large $n$, with probability at least $1-e^{-\delta n}$ in $m$
and $\lambda$, we have
\[
\norm{ P_m - N }_1 < e^{-\delta n} \enspace .
\]
\end{theorem}

The proof strategy is to bound $\Var_m \norm{\Pi_m \vb}^2$ using
Lemma~\ref{lem:var}, and apply Chebyshev's inequality to conclude that
it is almost certainly close to its expectation. Recall, however, that
our bounds on the variance of $\norm{\Pi_m \vb}^2$ depend on the
decomposition of $S^\lambda \otimes (S^\lambda)^*$ is into irreducibles and,
furthermore, on the projection of $\vb \otimes \vb^*$ into these
irreducible subspaces. Matters are somewhat complicated by the fact
that certain $S^\mu$ appearing in $S^\lambda \otimes (S^\lambda)^*$ may contribute
more to the variance than others. While Theorem~\ref{thm:roichman}
allows us to bound the contribution of those constituent irreducible
representations $S^\mu$ for which $\mu_1$ and $\mu'_1$ are much smaller
than $n$, those which violate this condition could conceivably
contribute large terms to the variance estimates.  Fortunately, in
this single coset case, the total fraction of the space $S^\lambda \otimes
(S^\lambda)^*$, dimensionwise, consisting of such $S^\mu$ is small with
overwhelming probability.  Despite this, we cannot preclude the
possibility that for a \emph{specific} vector $\vb$, the quantity
$\Var \norm{\Pi_m \vb}^2$ is large, as $\vb$ may project solely into
spaces of the type described above. On the other hand, as these
troublesome spaces amount to a small fraction of $S^\lambda \otimes (S^\lambda)^*$,
only a few $\vb$ can have this property.

Specifically, let $0 < c < 1/4$ be a constant, and let $\Lambda = \Lambda_c$
denote the collection of Young diagrams $\mu$ with the property that
either $\mu_1 \geq (1-c) n$ or $\mu_1' \geq (1-c) n$.  Then Part I establishes the following
upper bounds on the cardinality of $\Lambda$ and the dimension of any
$S^\mu$ with $\mu \in \Lambda$:
\begin{lemma}  \label{lem:lambda}
  Let $p(n)$ denote the number of integer partitions of $n$.  Then
  $|\Lambda| \leq 2 cn p(cn)$, and $d^\mu < n^{cn}$ for any $\mu \in \Lambda$.
\end{lemma}

As a result, the representations associated with diagrams in $\Lambda$
constitute a negligible fraction of $\widehat{S_n}$; specifically,
from Lemma~\ref{lem:lower}, part~\ref{item:lower-linear}, the
probability that a $\lambda$ drawn according to the Plancherel distribution
falls into $\Lambda$ is $n^{-\Omega(n)}$.
The following lemma shows that this is also true for the distribution
$P(\rho)$ induced on $\widehat{S_n}$ by weak Fourier sampling the coset state
$\ket{H}$.
\begin{lemma}  
\label{lem:notinlambda}
Let $d < 1/2$ be a constant and let $n$ be sufficiently large.  Then
there is a constant $\gamma > 0$ such that
we observe a representation $S^\lambda$ with $d^\lambda \geq 
n^{dn}$ with probability at least $1-n^{-\gamma n}$.
\end{lemma}

On the other hand, for a representation $S^\mu$ with $\mu \notin \Lambda$,
Theorem~\ref{thm:roichman} implies that
\begin{equation}
  \label{eq:small}
  \abs{\frac{\chi^\mu(M)}{d^\mu}}
  \leq \bigl( \max(q,1-c) \bigr)^{bn} \leq e^{-\alpha n} \enspace
\end{equation}
for a constant $\alpha \geq bc > 0$.  Thus the contribution of such an
irreducible to the variance estimate of Lemma~\ref{lem:var} is
exponentially small.  
The remainder of the proof of Theorem~\ref{thm:one-frames} uses a combination of 
Chebyshev's and Markov's inequalities to bound the total variation distance 
between $P_m$ and the natural distribution.

\section{Variance and decomposition for multiregister experiments}
\label{sec:multi}

We turn now to the multi-register case, where Steps 1, 2 and 3 are
carried out on $k$ independent registers.  This yields a state in
$\C[G^k]$, i.e.,
\[ \ket{c_1 H} \otimes \cdots \otimes \ket{c_k H} \]
where the $c_i$ are uniformly random coset representatives.  The
symmetry argument of Section 3 of Part I applies to each register, so
that the optimal measurement is consistent with first measuring the
representation name in each register.  However, the optimal
measurement generally does not consist of $k$ independent measurements
on this tensor product state; rather, it is entangled, consisting of
measurement in a basis whose basis vectors $\vb$ are not of the form
$\vb_1 \otimes \cdots \otimes \vb_k$.  As mentioned above, for the dihedral groups
in particular, a fair amount is known: Ip~\cite{Ip} showed that the
optimal measurement for two registers is entangled,
Kuperberg~\cite{Kuperberg03} showed that an entangled measurement
yields a subexponential-time algorithm for the hidden subgroup
problem, and Bacon, Childs and van Dam~\cite{BaconCvD} have calculated
the optimal measurement on $k$ registers.

Extending the results of part I to this case involving multiple coset
states will proceed in three steps:
\begin{itemize}
\item In Section~\ref{sec:multi-var}, we generalize the expectation
  and variance bounds of Lemma~\ref{lem:var} to the algebra $\C[G^k]$,
  viewed as a representation of $G$.
\item As in the single register proof, we must control the
  decomposition of the representations that appear in the expressions
  for expectation and variance. Unfortunately, the naive bounds
  applied in part I (relying on the fact that $\langle \chi_\sigma, \chi_\rho
  \chi_\tau\rangle_G \leq d_\sigma$ for irreducible representations $\rho$, $\sigma$, and
  $\tau$) are insufficient for controlling these decompositions. In
  Section~\ref{sec:multi-decomp}, we show how to bound the
  decomposition of these representations on average.
\item Finally, in Section~\ref{sec:two}, we show how to apply these
  results to eliminate the possibility of solving the HSP over $S_n$
  with any polynomial number of two-register experiments on coset
  states.
\end{itemize}

\subsection{Variance for Fourier sampling product states}
\label{sec:multi-var}

We begin by generalizing Lemmas~1, 2, and 3
of Part I to the multi-register case.  The reasoning is analogous
to that of Section 4 of Part I; the principal difficulty
is notational, and we ask the reader to bear with us. 

We assume we have measured the representation name on each of the
registers, and that we are currently in an irreducible representation
of $G^k$ labeled by $\rho_1 \otimes \cdots \otimes \rho_k$.  
Given a subset $I \subseteq \{1,\ldots,k\}$, we
can separate this tensor product into the registers inside $I$ and
those outside, and then decompose the product of those inside $I$ into
irreducibles $\sigma$:
\begin{eqnarray*}
\rho_1 \otimes \cdots \otimes \rho_k
& = & \bigotimes_{i \in I} \rho_i \otimes \bigotimes_{i \notin I} \rho_i \\
& = & \left( \bigoplus_{\sigma \prec \otimes_{i \in I} \rho_i} 
  a^I_\sigma \sigma \right) \otimes \bigotimes_{i \notin I} \rho_i 
\end{eqnarray*}
where $a^I_\sigma$ is the multiplicity of $\sigma$ in $\otimes_{i \in I} \rho_i$.  Now given an irrep $\sigma$, let $\Pi^I_\sigma$ denote the projection operator onto the subspace acted on by
\[ 
a^I_\sigma \sigma \otimes \bigotimes_{i \notin I} \rho_i \enspace .
\]
In other words, $\Pi^I_\sigma$ projects the registers in $I$ onto the subspaces isomorphic to $\sigma$, and leaves the registers outside $I$ untouched.  Note that in the case where $I$ is a singleton we have $\Pi^{\{i\}}_{\rho_i} = \one$.  

As before, the hidden subgroup is $H=\{1,m\}$ for an involution $m$
chosen at random from a conjugacy class $M$.  However, we now
have, in effect, the subgroup $H^k \subset G^k$, and summing over the
elements of $H^k$ gives the projection operator $\Pi_{H^k} = \Pi_m^{\otimes
  k}$.  The probability we observe an (arbitrarily entangled) basis
vector $\vb \in \rho_1 \otimes \cdots \otimes \rho_k$ is then \begin{equation}
\label{eq:pbk}
P_m(\vb) = \frac{\norm{\Pi_m^{\otimes k} \vb}^2}{\rank \Pi_m^{\otimes k}} \enspace . 
\end{equation}
When we calculate the expectation of this over $m$, we will find ourselves summing the following quantity over the subsets $I \subseteq \{1, \ldots, k\}$:
\begin{equation}
\label{eq:ei}
 E^I(\vb) = \sum_{\sigma \prec \otimes_{i \in I} \rho_i} 
\frac{\chi^\sigma(M)}{\dim \sigma} \norm{\Pi^I_\sigma \vb}^2 
\end{equation}
with $E^\emptyset(\vb) = \norm{\vb}^2$ (since an empty tensor product gives the trivial representation).

For the variance, we will find ourselves dealing with pairs of subsets $I_1, I_2 \subseteq \{1, \ldots, k\}$ and decompositions of the form
\begin{eqnarray*}
\left( \rho_1 \otimes \cdots \otimes \rho_k \right) \otimes
\left( \rho^*_1 \otimes \cdots \otimes \rho^*_k \right) 
& = & \left( \bigotimes_{i \in I_1} \rho_i \otimes \bigotimes_{i \in I_2} \rho^*_i \right) 
\otimes \left( \bigotimes_{i \notin I_1} \rho_i \otimes \bigotimes_{i \notin I_2} \rho^*_i \right) \\
& = & \left( \bigoplus_{\sigma \prec \bigotimes_{i \in I} \rho_i \otimes \bigotimes_{i \in I_2} \rho^*_i} 
  a^{I_1,I_2}_\sigma \sigma \right) 
\otimes \left( \bigotimes_{i \notin I_1} \rho_i \otimes \bigotimes_{i \notin I_2} \rho^*_i \right)
\end{eqnarray*}
just as we considered $\rho \otimes \rho^*$ in the one-register case.  We can then define a projection operator $\Pi^{I_1,I_2}_\sigma$ onto the subspace acted on by 
\[ a^{I_1,I_2}_\sigma \sigma 
\otimes \left( \bigotimes_{i \notin I_1} \rho_i \otimes \bigotimes_{i \notin I_2} \rho^*_i \right)
\]
and we define the following quantity, 
\begin{equation}
\label{eq:ei1i2}
E^{I_1,I_2}(\vb) = \sum_{\sigma \prec \bigotimes_{i \in I} \rho_i \otimes \bigotimes_{i \in I_2} \rho^*_i} 
\frac{\chi^\sigma(M)}{\dim \sigma} \norm{\Pi^{I_1,I_2}_\sigma (\vb \otimes \vb^*)}^2 
\end{equation}
with $E^{\emptyset,\emptyset}(\vb)=\norm{\vb}^4$.

We can now state the following lemma.  The reader can check
that~\eqref{eq:vark} corresponds exactly to Equation (4.3) of part I
in the one-register case.
\begin{lemma} \label{lem:var-k}
  Let $\vb \in \rho_1 \otimes \cdots \otimes \rho_k$ and let $m$ be an element chosen
  uniformly from a conjugacy class $M$ of involutions.  Then
\begin{eqnarray}
  \Exp_m \norm{\Pi_m^{\otimes k} \vb}^2 
  & = & \frac{1}{2^k} \left( 1 + \sum_{I \subseteq \{1,\ldots,k\} : I \neq \emptyset} 
  E^I(\vb) \label{eq:expk} \right) \\
   \Var_m \norm{\Pi_m^{\otimes k} \vb}^2 
  & = & \frac{1}{4^k} \sum_{I_1,I_2 \subseteq \{1,\ldots,k\} : I_1, I_2 \neq \emptyset} 
  E^{I_1,I_2}(\vb) - E^{I_1}(\vb) E^{I_2}(\vb)^* 
  \enspace .  
  \label{eq:vark}
\end{eqnarray}
\end{lemma}

\begin{proof}  Let $m^I$ denote the operator that operates on the $i$th register by $m$ for each $i \in I$ and leaves the other registers unchanged.  This acts on $\vb$ as $\tau(m)$ where $\tau = \bigotimes_{i \in I} \rho_i(m)$, and Lemma~\ref{lem:exp} implies that 
\[ \Exp_m \langle \vb, m^I \vb \rangle = E^I(\vb) \enspace . \]
Then~\eqref{eq:expk} follows from the observation that
\[ \Pi_m^{\otimes k} \vb = \frac{1}{2^k} \sum_{I \subseteq \{1,\ldots,k\} } m^I \vb \]
and so
\[ \Exp_m \norm{\Pi_m^{\otimes k} \vb}^2 
= \Exp_m \langle \vb, \Pi_m^{\otimes k} \vb \rangle
= \frac{1}{2^k} \sum_{I \subseteq \{1,\ldots,k\} } \Exp_m \langle \vb, m^I \vb \rangle 
= \frac{1}{2^k} \sum_{I \subseteq \{1,\ldots,k\} } E^I(\vb) \enspace . \]
Separating out the term $E^\emptyset(\vb) = \norm{\vb}^2$ completes the proof of~\eqref{eq:expk}.

Similarly, let the operator $m^{I_1} \otimes m^{I_2}$ act on $\vb \otimes \vb^*$ by multiplying the $i$th register of $\vb$ by $m$ whenever $i \in I_1$, and multiplying the $i$th register of $\vb^*$ whenever $i \in I_2$.  Then it acts as $\tau(m)$ where $\tau = \bigotimes_{i \in I_1} \rho_i(m) \otimes \bigotimes_{i \in I_2} \rho^*_i(m)$, and Lemma~\ref{lem:exp} implies 
\[ \Exp_m \langle \vb \otimes \vb^*, (m^{I_1} \otimes m^{I_2})(\vb \otimes \vb^*) \rangle 
= E^{I_1,I_2}(\vb) \enspace . \]
Then analogous to Lemmas~\ref{lem:second} and~\ref{lem:var}, the second moment is
\begin{eqnarray*}
\Exp_m \norm{\Pi_m^{\otimes k} \vb}^4
& = & \Exp_m \langle \vb, \Pi_m^{\otimes k} \vb \rangle \langle \vb^*, \Pi_m^{\otimes k} \vb^* \rangle \\
& = & \Exp_m \langle \vb \otimes \vb^*, (\Pi_m^{\otimes k} \otimes \Pi_m^{\otimes k}) (\vb \otimes \vb^*) \rangle \\
& = & \frac{1}{4^n} \sum_{I_1,I_2 \subseteq \{1,\ldots,k\} }
\Exp_m \langle \vb \otimes \vb^*, (m^{I_1} \otimes m^{I_2}) (\vb \otimes \vb^*) \rangle \\
& = & \frac{1}{4^n} \sum_{I_1,I_2 \subseteq \{1,\ldots,k\} } E^{I_1,I_2}(\vb)
\end{eqnarray*}
and so the variance is
\begin{eqnarray*}
 \Var_m \norm{\Pi_m^{\otimes k} \vb}^2 
  & = & \Exp_m \norm{\Pi_m^{\otimes k} \vb}^4 
  - \left( \Exp_m \norm{\Pi_m^{\otimes k} \vb}^2 \right)^2 \\
  & = & \frac{1}{4^k} \sum_{I_1,I_2 \subseteq \{1,\ldots,k\} } 
  E^{I_1,I_2}(\vb) - E^{I_1}(\vb) E^{I_2}(\vb)^* 
  \enspace .
\end{eqnarray*}
Finally,~\eqref{eq:vark} follows from the fact that the two terms in the sum cancel whenever $I_1$ or $I_2$ is empty.
\end{proof}

\subsection{The associated Clebsch-Gordan problem}
\label{sec:multi-decomp}

The expressions $E^I(\vb)$ and $E^{I_1,I_2}(\vb)$ above depend on the
decomposition of representations of the form
\[  \bigotimes_{i \in I_1} \rho_i \otimes \bigotimes_{i \in I_2} \rho^*_i 
= \left( \bigotimes_{i \in I_1 \setminus I_2} \rho_i \otimes
  \bigotimes_{i \in I_2 \setminus I_1} \rho^*_i \right) \otimes
\bigotimes_{i \in I_1 \cap I_2} (\rho_i \otimes \rho^*_i)\enspace.
\]
Moreover, since the Plancherel distribution is symmetric with respect
to conjugation, this is a tensor product of $\abs{I_1 \triangle I_2}$
representations $\rho_i$ with $\abs{I_1 \cap I_2}$ representations $\sigma_j
\otimes \sigma^*_j$, where both the $\rho_i$ and the $\sigma_j$ are chosen according
to the Plancherel distribution.  This motivates the following
definition.

\begin{definition}
  \label{def:v}
  For non-negative integers $k$ and $\ell$ and $\vrho = (\rho_1, \ldots, \rho_k) \in
  \widehat{G}^k$ and $\vsigma = (\sigma_1,\ldots, \sigma_\ell) \in \widehat{G}^\ell$, let
  $V(\vrho, \vsigma)$ denote the representation
  $$
  \bigotimes_{i=1}^k \rho_i \otimes \bigotimes_{j = 1}^\ell (\sigma_j \otimes \sigma_j^*)\enspace.
  $$
\end{definition}

Of particular interest is the dimensionwise fraction of such
representations consisting of low-dimensional irreducibles. For these
representations, the naive decomposition results
of Equation (6.4) of Part I
no longer suffice to obtain nontrivial
estimates.  Fortunately, the combinatorial representations $\reg$ and
$\conj$ discussed in Section~\ref{sec:representation-theory} can be
used to control the structural properties of these tensor products
\emph{on average}. We will apply this machinery in
Section~\ref{sec:two} to control general two-register experiments.

Recall that the multiplicity of an irreducible representation $\tau$ in the decomposition of
a representation $V$ into irreducibles is the inner product $\chlangle \chi_\tau, \chi_V
\chrangle{G}$, and that $[g]$ denotes the conjugacy class of $g$.

\begin{lemma}
  \label{lem:average-multiplicity}
  Fix $\tau \in \widehat{G}$ and let $\vrho$ and $\vsigma$ be random
  variables taking values in $\widehat{G}^k$ and $\widehat{G}^\ell$,
  respectively, so that each $\rho_i$ and $\sigma_j$ is independently
  distributed according to the Plancherel distribution.  Then
\begin{eqnarray*}
  \Exp_{\vrho,\vsigma} \frac{\left\chlangle \chi_\tau,\chi_{V(\vrho,\vsigma)}\right\chrangle{G}}
    {\dim V(\vrho,\vsigma)}
& = & \frac{d_\tau}{|G|} \quad \text{if}\;k > 0 \enspace , \\
  \Exp_{\vrho,\vsigma} \frac{\left\chlangle \chi_\tau,\chi_{V(\vrho,\vsigma)}\right\chrangle{G}}
    {\dim V(\vrho,\vsigma)}
    & \leq & \frac{d_\tau}{|G|} \sum_g \frac{1}{| [g] |^{\ell}} \quad \text{if}\;k = 0 \enspace . 
\end{eqnarray*}
\end{lemma}

\begin{proof}
  The two permutation representations $\reg$ and $\conj$ will play a
  special role in the analysis: in particular, we will see that the
  expectation of interest can be expressed in terms of these
  combinatorial characters. Specifically,
  \begin{align}
    \Exp_{\vrho,\vsigma}& \left[\frac{\left\chlangle \chi_\tau,\chi_{V(\vrho,\vsigma)}\right\chrangle{G}}{\dim
        V(\vrho,\vsigma)}\right] = \sum_{\vrho \in \widehat{G}^k}
    \sum_{\vsigma \in \widehat{G}^\ell} \left(\prod_i \frac{d_{\rho_i}^2}{|G|}\right)
    \left(\prod_j \frac{d_{\sigma_j}^2}{|G|}\right) \frac{\chlangle \chi_\tau,
      \chi_{V(\vrho,\vsigma)}\chrangle{G}}{\dim V(\vrho,\vsigma)}\\ 
    \label{eqn:applying-dimension} &= \frac{1}{|G|^{k+\ell}} \sum_{\vrho \in
      \widehat{G}^k} \sum_{\vsigma \in \widehat{G}^\ell} \left(\prod_i
      d_{\rho_i}\right) \chlangle \chi_\tau, \chi_{V(\vrho,\vsigma)}\chrangle{G}\\ 
    \label{eqn:applying-character} &= \frac{1}{|G|^{k+\ell}} \left\chlangle \chi_\tau ,
      \sum_{\vrho \in \widehat{G}^k} \sum_{\vsigma \in
        \widehat{G}^\ell} \left(\prod_i d_{\rho_i} \chi_{\rho_i}\right) \left( \prod_j \chi_{\sigma_j} \chi_{\sigma_j}^*\right)\right\chrangle{G}\\
    &= \frac{1}{|G|^{k+\ell}} \left\chlangle \chi_\tau , \chi_\reg^k \chi_\conj^\ell \right\chrangle{G}
    \end{align}
    where the equality of line~\eqref{eqn:applying-dimension} follows
    from the fact that the dimension of $V(\vrho,\vsigma)$ is
    $\prod_i d_{\sigma_i} \cdot \prod_j d_{\sigma_j}^2$ and the equality of
    line~\eqref{eqn:applying-character} follows from the fact that the
    character of $V(\vrho,\vsigma)$ is $\prod_i \chi_{\rho_i} \prod_j
    \chi_{\sigma_j}\chi_{\sigma_j}^*$.
    
    Recall that for any representation $\upsilon$ we have $\chi_\upsilon(1) = d_\upsilon$. As
    $\chi_\reg$ is a multiple of the delta function $\delta_g$, whenever $k > 1$ we have 
    $\chlangle\chi_\tau , \chi_\reg^k \chi_\conj^\ell\chrangle{G} = d_\tau |G|^{k + \ell - 1}$ and 
    $$
    \Exp_{\vrho,\vsigma} \left[\frac{\left\chlangle \chi_\tau,\chi_{V(\vrho,\vsigma)}\right\chrangle{G}}{\dim
        V(\vrho,\vsigma)}\right] = \frac{d_\tau}{|G|}\enspace.
    $$
    On the other hand, when $k = 0$ we have
    \begin{align*}
      \Exp_{\vrho,\vsigma} & \left[\frac{\left\chlangle \chi_\tau,\chi_{V(\vrho,\vsigma)}\right\chrangle{G}}{\dim
        V(\vrho,\vsigma)}\right] = \frac{1}{|G|^\ell} \left\chlangle \chi_\tau , \chi_\conj^\ell \right\chrangle{G} = \frac{1}{|G|^{\ell+1}} \sum_g \chi_\tau^*(g) \chi_\conj^\ell(g)\\
      &= \frac{1}{|G|^{\ell+1}} \sum_g \chi_\tau^*(g) \frac{|G|^\ell}{\abs{[g]}^\ell } = \frac{1}{|G|} \sum_g \chi_\tau^*(g) \frac{1}{\abs{[g]}^\ell }\\
      & \leq \frac{d_\tau}{|G|} \sum_g \frac{1}{\abs{[g]}^\ell } \enspace , \\
    \end{align*}
    where the last inequality follows from the fact that $|\chi_\tau(g)|
    \leq d_\tau$ for all $g$.
\end{proof}

Now note that the sum $\sum_g 1/\abs{[g]}^\ell$ can also be written as a sum over the conjugacy classes $C$.  In particular, if $\ell \geq 2$, 
\[ \sum_g \frac{1}{\abs{[g]}^\ell} = \sum_C \frac{1}{|C|^{\ell-1}} \leq \sum_C \frac{1}{|C|} \enspace . \] 
In the case of the symmetric group $S_n$, the next lemma shows that
this quantity is in fact $1 + o(1)$.

\begin{lemma}
  Given a partition $\lambda = (\lambda_1, \ldots, \lambda_t)$ with $\sum_i \lambda_i = n$ and
  $\lambda_i \geq \lambda_{i+1}$ for all $i$, let $C_\lambda$ denote the conjugacy
  class of $S_n$ consisting of permutations with cycle structure $\lambda$.
  Then
\[ \sum_\lambda \frac{1}{|C_\lambda|} = 1+o(1) \enspace . \]
\end{lemma}

\begin{proof}  First note that if we group the $\lambda_i$ into blocks
  consisting of $\tau_1$ $1$s, $\tau_2$ $2$s, and so on (such that $\sum_i
  \tau_i i = n$) then the size of the conjugacy class is given by
\[ |C_\lambda| = \frac{n!}{\left( \prod_i \tau_i! \right) \left( \prod_i \lambda_i \right)} \]
since we can cyclically permute the elements of each cycle, and permute cycles of the same size with each other.  Thus 
\begin{equation}
\label{eq:conjsum}
 \sum_\lambda \frac{1}{|C_\lambda|}  
=\sum_\lambda  \frac{1}{n!} \left( \prod_i \tau_i! \right) \left( \prod_i \lambda_i \right)  \enspace . 
\end{equation}
Now suppose that the conjugacy class consists of elements with support $s$, i.e., with $\tau_1 = n-s$ fixed points.  Since we can specify such a partition with a partition of $s$ objects, the number of such partitions is at most $p(s)$.  Moreover we have
\[ \prod_i \tau_i! = (n-s)! \prod_{i \geq 2} \tau_i! \leq (n-s)! \,(s/2)! \]
and
\[ \prod_i \lambda_i = \prod_{\lambda > 1} \lambda \leq e^{s/e} \enspace . \]
since this is true for any set of reals $\lambda \geq 0$ such that $\sum \lambda =
s$.  Then~\eqref{eq:conjsum} becomes
\begin{equation}
\label{eq:conjsum2}
 \sum_\lambda \frac{1}{|C_\lambda|}  
\leq \sum_{s=0}^n \frac{(n-s)! \,(s/2)!}{n!} \,p(s) e^{s/e} 
\end{equation}
Now, for $s > \sqrt{n}$, we have
\begin{equation}
\label{eq:conjbound}
\frac{(n-s)! \,(s/2)!}{n!} 
= \frac{1}{\binom{n}{s}} \frac{(s/2)!}{s!} 
\leq \frac{(s/2)!}{s!} 
< \left( \frac{2s}{e} \right)^{-s/2} 
\leq n^{-s/4} 
\end{equation}
and for $s \leq \sqrt{n}$, for sufficiently large $n$ a stronger bound
holds,
\[ \frac{(n-s)! \,(s/2)!}{n!} 
\leq \frac{(\sqrt{n}/2)^{s/2}}{(n-s)^s} \leq n^{-3s/4} \enspace . \]
Thus~\eqref{eq:conjbound} holds for all $s$.  Using the absurdly crude
bound $p(s) \,e^{s/e} < 4^s$, ~\eqref{eq:conjsum2} then becomes
\[ 
\sum_\lambda \frac{1}{|C_\lambda|} \leq \sum_{s=0}^n n^{-s/4} 4^s <
\frac{1}{1-4n^{-1/4}} = 1+O(n^{-1/4}) \enspace .
\]
\end{proof}

On the other hand, if $\ell = 1$ then the sum $\sum_g 1/\abs{[g]}$ is
simply the number of conjugacy classes.  Therefore, in the case of the
symmetric group, we have the following corollary of
Lemma~\ref{lem:average-multiplicity}.
\begin{corollary}
  \label{cor:sn-average-multiplicity}
  Let $G = S_n$ and $k$, $\ell$, $\rho$, and $\sigma$ as in
  Lemma~\ref{lem:average-multiplicity}. Then for any irreducible character $\chi_\tau$, 
  $$
  \Exp_{\vrho,\vsigma} \frac{\left\chlangle
      \chi_\tau,\chi_{V(\vrho,\vsigma)}\right\chrangle{S_n}} {\dim
    V(\vrho,\vsigma)} \leq (1 + o(1)) \,\frac{d_\tau}{n!}
  $$
  unless $k=0$ and $\ell = 1$, in which case
  $$
  \Exp_{\vrho,\vsigma} \frac{\left\chlangle \chi_\tau,\chi_{V(\vrho,\vsigma)}\right\chrangle{S_n}}
  {\dim V(\vrho,\vsigma)}
  \leq \frac{d_\tau p(n)}{n!} \enspace.
  $$
\end{corollary}

\subsection{A transverse projection lemma}

As a final technical preparatory element, we record a projection lemma
concerning the relationship between tensor powers of bases and
``transverse'' subspaces.

\begin{lemma}
\label{lem:proj}
  Let $U$ and $Y$ be finite-dimensional Hilbert spaces and let $B$ be
  an orthonormal basis for $U \otimes Y$. Let $W$ be a subspace of $U \otimes U^*$
  and $\Pi_W$ be the projection operator onto $W$. Then
  $$
  \sum_{\vb \in B} \abs{ (\Pi_W \otimes \one_{Y \otimes Y^*}) (\vb \otimes \vb^*) }^2 
  \leq \dim Y \times \dim W 
  \enspace . 
  $$
\end{lemma}

\begin{proof}
  Since $\Pi_W$ can be written as a sum of one-dimensional projection operators, 
  it suffices to establish the lemma in the case where $\dim W = 1$.
  Let $\{ u_i \}$ be an orthonormal basis for $U$.  Then we can write $\Pi_W = \Pi_w$ 
  where
  \[ w = \sum_{i,j} a_{ij} (u_i \otimes u_j^*) \enspace . \]
  Now let $\{ z_k \}$ be an orthonormal basis for $Y$.  Then 
  \begin{align*}
    \abs{(\Pi_w \otimes \one_{Y \otimes Y^*})(\vb \otimes \vb^*)}^2 
    &= \sum_{k,\ell} \abs{\inner{w \otimes (z_k \otimes z_\ell^*)}{\vb \otimes \vb^*}}^2 \\
    &= \sum_{k,\ell} \abs{ \sum_{i,j} a_{i,j} 
    \inner{(u_i \otimes u_j^*) \otimes (z_k \otimes z_\ell^*)}{\vb \otimes \vb^*}}^2 \\
    &= \sum_{k,\ell} \abs{ \sum_{i,j} a_{i,j} 
    \inner{u_i \otimes z_k}{\vb} \inner{u_j^* \otimes z_\ell^*}{\vb^*}}^2 \\
    &\le \sum_{k,\ell} \left( \sum_i \abs{\inner{u_i \otimes z_k}{\vb}}^2 \right) 
    \left( \sum_i \abs{\sum_j a_{i,j} \inner{u_j^* \otimes z_\ell^*}{\vb^*}}^2 \right) \\
    &=  \sum_{i,\ell} \abs{\inner{\sum_j a_{i,j} u_j^* \otimes z_\ell^*}{\vb^*}}^2
  \end{align*}
  where we used the Cauchy-Schwartz inequality in the fourth line.  
  Summing over all $\vb \in B$ then gives
  \begin{align*}
  \sum_{\vb \in B} & \abs{ (\Pi_w \otimes \one_{Y \otimes Y^*}) (\vb \otimes \vb^*) }^2 
    \leq \sum_{i,\ell} \abs{\sum_j a_{i,j} u_j^* \otimes z_\ell^*}^2 \\
    & = \sum_{i,j} \abs{a_{ij}}^2 \dim Y 
     = \dim Y \enspace,
  \end{align*}
  as desired.
\end{proof}

\section{Two registers are insufficient for the symmetric group}
\label{sec:two}

In this section we show that no polynomial number of two-register
experiments can distinguish the involutions we have been considering
in $S_n$ from each other or from the trivial subgroup.  As in
Section~\ref{sec:multi}, we assume we have measured the representation
name on each of the two registers, and that we observed the
irreducible representations $S^\lambda$ and $S^\mu$.  For simplicity we
present the proof for von Neumann measurements; the generalization to
arbitrary frames $\{ \vb \}$ proceeds exactly as in the proof of
Theorem~\ref{thm:one-frames}.

\begin{theorem}
\label{thm:two}
Let $B=\{\vb\}$ be an orthonormal basis for $S^\lambda \otimes S^\mu$.
Given the hidden subgroup $H=\{1,m\}$ where $m$ is chosen uniformly at random from $M$, 
let $P_m(\vb)$ be the
probability that we observe the vector $\vb$ conditioned on having observed the representation names $S^\lambda$ and $S^\mu$, and let $U$ be the uniform distribution on $B$.  Then there
is a constant $\delta > 0$ such that for sufficiently large $n$, with probability at least
$1-e^{-\delta \sqrt{n}/\log n}$ in $m$, $\lambda$ and $\mu$, we have
\[ \norm{ P_m - U }_1 < e^{-\delta \sqrt{n} / \log n} \enspace . \]
\end{theorem}

\begin{proof}
  For $k=2$, Lemma~\ref{lem:var-k} specializes to the following:
\begin{eqnarray}
  \Exp_m \norm{\Pi_m^{\otimes 2} \vb}^2 
  & = & \frac{1}{4} \left( 1 + \sum_{I \subseteq \{\lambda,\mu\} : I \neq \emptyset} 
  E^I(\vb) \label{eq:exp2} \right) \\
  \Var_m \norm{\Pi_m^{\otimes 2} \vb}^2 
  & = & \frac{1}{16} \sum_{I_1,I_2 \subseteq \{\lambda,\mu\} : I_1, I_2 \neq \emptyset} 
  E^{I_1,I_2}(\vb) - E^{I_1}(\vb) E^{I_2}(\vb)^* 
  \enspace .  
  \label{eq:var2}
\end{eqnarray}
As before, $S^\lambda$ and $S^\mu$ are chosen with the distribution $P(\rho)$. 
Since this is exponentially close to the Plancherel distribution~\cite{HallgrenRT00}, 
we can use Lemma~\ref{lem:average-multiplicity} to calculate the expectations 
over $\lambda$ and $\mu$ of $E^I(\vb)$ and $E^{I_1,I_2}(\vb)$ with negligible error.  
We will then show that $E^I(\vb)$ and $E^{I_1,I_2}(\vb)$ are
superpolynomially small with the stated probability, for all but a
small fraction of basis vectors $\vb$, namely those that project into
low-dimensional representations.  As in Theorem~\ref{thm:one-frames}, we will then
use Markov's inequality to control the number of these basis vectors
and use Chebyshev's inequality to control the rest, and thus bound the 
total distance $\norm{ P_m - U }_1$.

However, the analysis, at least when $|I_1|=|I_2|= 2$, is more
delicate than for the one-register case.  As before, we exclude a set
of low-dimensional representations $\Lambda$, but now we restrict $\Lambda$ to
Young diagrams with width or height extremely close to $n$.
Specifically, let $c > 0$ be a constant to be determined below, and let
$\Lambda = \Lambda_c$ be the set of Young diagrams $\nu$ such that
\[ \max(\nu_1, \nu'_1) \geq n-c \sqrt{n}/\ln n \enspace . \]
Analogously to~\eqref{eq:small}, 
Theorem~\ref{thm:roichman} provides
the following bound on the characters $\chi^\nu$ for $\nu \notin \Lambda$,
\begin{equation}
  \label{eq:small2}
  \abs{\frac{\chi^\nu(M)}{d^\nu}}
  \leq \left(1-\frac{c}{\sqrt{n} \ln n} \right)^{bn} < e^{-\alpha \sqrt{n}/\ln n}
\end{equation}
where $\alpha = bc > 0$.  The size and dimension of $\Lambda$ is bounded by the following lemma.
\begin{lemma}  
\label{lem:lambda2}  
    $|\Lambda| = e^{o(\sqrt{n})}$ and $d^\nu < e^{c\sqrt{n}}$ for any $\nu \in \Lambda$.  Therefore, $\sum_{\nu \in \Lambda} (d^\nu)^2 < e^{2c\sqrt{n} + o(\sqrt{n})}$.
\end{lemma}

\begin{proof}  The proof of Lemma~\ref{lem:lambda} applies, except now $|\Lambda| < 2 x p(x)$ where $x=c \sqrt{n} / \ln n$.  
\end{proof}

Then the next lemma shows that with high probability in $\lambda$ and $\mu$, $E^I(\vb)$ is superpolynomially small for \emph{all} $\vb \in B$.  (Indeed, it is exponentially small for all but a few $\vb$, but we give this statement for simplicity.)
\begin{lemma}\label{lem:two-exp}
  Let $S^\mu$ and $S^\lambda$ be distributed according to the Plancherel
  distribution in $\widehat{S_n}$. Let $I \subseteq \{\lambda ,\mu\}$, $I \neq \emptyset$.  There is a constant $\gamma > 0$ such that for sufficiently large $n$, with probability $1-e^{-\gamma \sqrt{n}}$, $\abs{E^I(\vb)} \leq e^{-\alpha \sqrt{n}/\ln n}$ for all $\vb \in B$.
\end{lemma}

\begin{proof}  The case when $|I|= 1$ is identical to the one-register case, since then $E^I(\vb) = \chi^\lambda(M)/d^\lambda$.  Lemma~\ref{lem:lower} implies $\lambda \notin \Lambda$ with probability $1-e^{-\delta\sqrt{n}}$, and~\eqref{eq:small2} completes the proof of this case.
  
  For the case $|I| = 2$, it suffices to ensure that $S^\lambda \otimes S^\mu$
  contains no low-dimensional representations.  Let $\nu \in \Lambda$; then by
  Lemma~\ref{lem:average-multiplicity} and Lemma~\ref{lem:lambda2},
  the expected multiplicity of $S^\nu$ in $S^\lambda \otimes S^\mu$ is
\[ \Exp_{\lambda,\mu} \chlangle \chi^\nu, \chi^\lambda \chi^\mu \chrangle{S_n} 
= \Exp_{\lambda,\mu} d^\lambda d^\mu 
   \frac{\chlangle \chi^\nu, \chi^\lambda \chi^\mu \chrangle{S_n}}{d^\lambda d^\mu}
\leq e^{-2\hat{c}\sqrt{n}} d^\nu \leq e^{(c-2\hat{c})\sqrt{n}}
\]
where $\hat{c}$ is the constant appearing in
Theorem~\ref{thm:max-dimension}.  Thus if $c < \hat{c}$,
Lemma~\ref{lem:lambda2} and Markov's inequality imply that the
probability \emph{any} $S^\nu$ with $\nu \in \Lambda$ appears in $S^\lambda \otimes S^\mu$ is at
most $e^{-\hat{c}\sqrt{n}}$.  If none do, then~\eqref{eq:small2} and
the fact that $\abs{E^I} \leq \max_{\nu \notin \Lambda} \abs{\chi^\nu(M)/d^\nu}$ complete the
proof with $\gamma = \hat{c}$.
\end{proof}

For the variance estimates, for each $S^\lambda, S^\mu \in \widehat{S_n}$ and
$I_1, I_2 \subset \{\lambda,\mu\}$, recall Definition~\ref{def:v} and let
$$
V[I_1, I_2] = V(I_1 \triangle I_2, I_1 \cap I_2)\enspace,
$$
where $I_1 \triangle I_2$ is the symmetric difference. (We abuse notation
here, allowing, e.g., the set $I_1 \triangle I_2$ to stand for the tuple
of representations $S^\lambda$ with $\lambda \in I_1 \triangle I_2$.)  
For the variance calculation, as in the
single-register case, let $L[I_1,I_2] \subset V[I_1,I_2]$ be the subspace
consisting of copies of representations $S^\nu$ with $\nu \in \Lambda$, and let
$\Pi_{L[I_1,I_2]}$ be the projection operator onto this subspace; 
note that the projection oerator in~\eqref{eq:ei1i2} is $\Pi_{L[I_1,I_2]} \otimes \one$, 
where $\one$ acts on $V[\overline{I_1},\overline{I_2}]$. 
We will abbreviate $L = L[I_1,I_2]$ and $V = V[I_1,I_2]$ when the
parameters are clear from context. Then the following lemma bounds the
dimension of this subspace.
\begin{lemma}\label{lem:two-decomp}
  Let $S^\lambda$ and $S^\mu$ be distributed according to the Plancherel
  distribution in $\widehat{S_n}$ and let $I_1, I_2 \subseteq \{\lambda,\mu\}$,
  $I_1, I_2 \neq \emptyset$.  There is a constant $\beta > 0$ such that for
  sufficiently large $n$, with probability at least $1-e^{-\beta\sqrt{n}}$, 
 \[ \sum_{\vb \in B} (\Pi_{L[I_1,I_2]} \otimes \one)(\vb \otimes \vb^*) \leq e^{-\beta \sqrt{n}} |B| \enspace . \]
\end{lemma}

\begin{proof}  
  If $|I_1|=|I_2|=1$ and $I_1 \neq I_2$, then the proof of the previous
  lemma shows that $L$ is in fact empty with probability $1-e^{-\Omega(\sqrt{n})}$.  
  When $I_1 = I_2$ and $|I_1| = 1$, however, this is not true; taking $I_1=I_2=\{ \lambda \}$, 
  $S^\lambda \otimes (S^\lambda)^*$ contains exactly one copy of the
  trivial representation.  However, applying Lemma~\ref{lem:proj} 
  (with $U=S^\lambda$, $Y=S^\mu$, and $W=L$), 
  Corollary~\ref{cor:sn-average-multiplicity}, Theorem~\ref{thm:max-dimension}, 
  Lemma~\ref{lem:lambda2}, and~\eqref{eq:pn} gives
  \begin{align*}
    \Exp_{\lambda,\mu} & \frac{1}{|B|} \sum_{\vb \in B} (\Pi_L \otimes \one)(\vb \otimes \vb^*)
    = \Exp_{\lambda,\mu} \frac{d^\mu \dim L}{|B|} \\
    &= \Exp_{\lambda,\mu} \frac{1}{d^\lambda} \sum_{\nu \in \Lambda} d^\nu 
    \chlangle \chi^\nu, (\chi^\lambda)^2 \chrangle{S_n} \\
    & \le \Exp_{\lambda,\mu} \frac{d^\lambda p(n)}{n!} \sum_{\nu \in \Lambda} (d^\nu)^2 \\
    & \leq \frac{e^{(-\hat{c} + \delta + 2c) \sqrt{n} + o(\sqrt{n})}}{\sqrt{n!}}
    = n^{-\Omega(n)}
  \end{align*}
where we recall that $|B|=d^\mu d^\lambda$ and $\dim V=(d^\lambda)^2$.
  
  When $|I_1| = 2$ and $|I_2| = 1$, e.g. $I_1=\{\lambda,\mu\}$ and $I_2=\{\lambda\}$, 
  then $\dim V = (d^\lambda)^2 d^\mu$.  Applying Lemma~\ref{lem:average-multiplicity} and 
  taking into account the fact that $\one$ acts on a space of dimension $d^\mu$, we have
  \begin{align*}
    \Exp_{\lambda,\mu} & \frac{1}{|B|} \sum_{\vb \in B} (\Pi_L \otimes \one)(\vb \otimes \vb^*)
    = \Exp_{\lambda,\mu} \frac{d^\mu \dim L}{|B|} \\
    &= \Exp_{\lambda,\mu} \frac{1}{d^\lambda} \sum_{\nu \in \Lambda} d^\nu 
    \chlangle \chi^\nu, (\chi^\lambda)^2 \chi^\mu \chrangle{S_n} \\
    & \leq \Exp_{\lambda,\mu} \frac{d^\lambda d^\mu}{n!} \sum_{\nu \in \Lambda} (d^\nu)^2 \\
    & \leq e^{(-2\hat{c}+2c)\sqrt{n} + o(\sqrt{n})} 
    < e^{-\hat{c}\sqrt{n}} 
  \end{align*}
  if we set $c < \hat{c}/2$.  The case when $|I_2| = 2$ and $|I_1| = 1$ is identical.
  
  Finally, we consider the case when $|I_1|=|I_2|= 2$.  Now
  $\dim V = (d^\lambda)^2 (d^\mu)^2$, and
  Corollary~\ref{cor:sn-average-multiplicity} gives
  \begin{align*}
  \Exp_{\lambda,\mu} & \frac{1}{|B|} \sum_{\vb \in B} (\Pi_L \otimes \one)(\vb \otimes \vb^*) 
    = \Exp_{\lambda,\mu} \frac{\dim L}{|B|} \\
    &= \Exp_{\lambda,\mu} \frac{1}{d^\lambda d^\mu} \sum_{\nu \in \Lambda} d^\nu 
    \chlangle \chi^\nu, (\chi^\lambda)^2 \chrangle{S_n} \\
    & \leq (1+o(1))  \Exp_{\lambda,\mu} \frac{d^\lambda d^\mu}{n!} \sum_{\nu \in \Lambda} (d^\nu)^2 
  \end{align*}
  which, as in the previous case, is less than $e^{-\hat{c}\sqrt{n}}$
  if we set $c < \hat{c}/2$.  
  
  Thus, in all three cases
  we have 
  \[ \Exp_{\lambda,\mu} \frac{1}{|B|} \sum_{\vb \in B} (\Pi_L \otimes \one)(\vb \otimes \vb^*) < e^{-\hat{c}\sqrt{n}} \enspace . \]
  By Markov's inequality, the
  probability that $\dim L > e^{-(\hat{c}/2)\sqrt{n}} |B|$ is at most
  $e^{-(\hat{c}/2)\sqrt{n}}$.  Thus setting $\beta < \hat{c}/2$
  completes the proof.
\end{proof}

Now, let $E_0$ denote the following event:
\begin{enumerate}
\item \label{item:conditioned-characters}
  $\max\left(|\chi^\lambda(M)/d^\lambda|,|\chi^\mu(M)/d^\mu|\right)
  \leq e^{-\Omega(n)}$,
\item \label{item:conditioned-single-set} $\abs{E^I(\vb)} = e^{-\alpha
    \sqrt{n}/ \ln n}$ for all $\vb \in B$ and all $I \subset
  \{\lambda,\mu\}$, and
\item \label{item:conditioned-small-fraction} $\sum_{\vb \in B} (\Pi_L \otimes \one)(\vb \otimes \vb^*) 
\leq e^{- \beta \sqrt{n}} |B|$ for each $I_1,I_2 \subset \{\lambda,\mu\}$
  with $I_1 \neq \emptyset$ and $I_2 \neq \emptyset$.
\end{enumerate}
As a consequence of~\eqref{eq:small} and Lemmas~\ref{lem:notinlambda},
\ref{lem:two-exp} and~\ref{lem:two-decomp}, $E_0$ occurs with
probability $1 - e^{-\Omega(\sqrt{n})}$.  In what follows we condition
on $E_0$.  This will allow us to control the three principal
parameters that determine the total variation distance between $P_m$
and the uniform distribution: $\rank \Pi_m^{\otimes 2}$,
$\Exp_m[\Pi_m^{\otimes 2}(\vb)]$, and $\Var_m[\Pi_m^{\otimes
  2}(\vb)]$.

Considering $\rank \Pi_m^{\otimes 2}$, note that the rank of
$\Pi_m^{\otimes 2}$ restricted to a representation $S^\lambda \otimes
S^\mu$ is the product of the ranks of $\Pi_m$ restricted to
$S^\lambda$ and $S^\mu$; then~\eqref{eq:rank}, and
item~\ref{item:conditioned-characters} of $E_0$ give
\begin{equation}
\label{eq:approx-rank2}
\rank \Pi_m^{\otimes 2} 
= \frac{d^\mu d^\lambda}{4}\left(1 + \frac{\chi^\mu(M)}{d^\mu}\right)\left(1 + \frac{\chi^\lambda(M)}{d^\lambda}\right)
= \frac{|B|}{4} \left( 1+e^{-\Omega(n)} \right)
\enspace.
\end{equation}

As for the expectation $\Exp_m[\Pi_m^{\otimes 2}(\vb)]$, in light
of~\eqref{eq:exp2} and item~\ref{item:conditioned-single-set} of
$E_0$, we conclude that for each $\vb \in B$,
\begin{equation}\label{eq:exp-2}
\left| \Exp_m \norm{\Pi_m^{\otimes 2} \vb}^2 - \frac{1}{4}\right| \leq 3
e^{-\alpha \sqrt{n}/ \ln n}\enspace.
\end{equation}

Finally, we focus on the variance. Define $B_L \subset B$ to be the
set of basis vectors $\vb$ such that for some nontrivial $I_1, I_2
\subset \{ \lambda, \mu\}$, $\norm{(\Pi_{L[I_1,I_2]} \otimes \one) (\vb \otimes \vb^*)}^2 \geq
e^{-(\beta/2) \sqrt{n}}$.  Then since
item~\ref{item:conditioned-small-fraction} of $E_0$ holds for each of
the $3^2=9$ pairs of nonempty subsets $I_1, I_2$, we have 
\[ |B_L|
\leq e^{(\beta/2) \sqrt{n}} \sum_{I_1,I_2} 
\sum_{\vb \in B} (\Pi_{L[I_1,I_2]} \otimes \one)(\vb \otimes \vb^*) 
\leq 9 e^{-(\beta/2) \sqrt{n}} |B| 
\enspace .  \]
Observe that for any $\vb
\in B \setminus B_L$, Equations~\eqref{eq:var2},~\eqref{eq:small2},
and item~\ref{item:conditioned-single-set} of $E_0$ give
\begin{equation}
\label{eq:good-var-2}
  \Var_m \norm{\Pi_m^{\otimes 2} \vb}^2 
  \leq \frac{9}{16} \left( e^{-\alpha \sqrt{n}/ \ln n} + e^{-2\alpha \sqrt{n}/ \ln n} + e^{-(\beta/2) \sqrt{n}} \right) 
  < e^{-\alpha \sqrt{n}/\ln n}
  \enspace.
\end{equation}
Then Chebyshev's inequality gives
\begin{equation} 
\label{eq:chebyshev2}
\Pr \left[ \,\abs{ \norm{\Pi_m^{\otimes 2} \vb}^2 - \Exp_m \norm{\Pi_m^{\otimes 2} \vb}^2 }
\geq e^{-(\alpha/3) \sqrt{n}/\ln n} \right] \leq e^{-(\alpha/3) \sqrt{n}/\ln n} \enspace .
\end{equation}
Analogous to Theorem~\ref{thm:one-frames}, let $\bad \subset B \setminus B_L$ 
denote the subset of basis vectors for which the event of~\eqref{eq:chebyshev2} is violated.
(As in the one-register case, while $B_L$ depends only on the choice of $\lambda$ and $\mu$, $\bad$ depends also on $m$.)  Let $E_1$ denote the event 
\[ |\bad| < e^{-(\alpha/6) \sqrt{n}/\ln n} |B| \enspace . \]
Then~\eqref{eq:chebyshev2} and Markov's inequality imply that $E_1$ occurs with probability $1-e^{-(\alpha/6) \sqrt{n}/\ln n}$.

So, finally, recall that $P_m(\vb) = \norm{\Pi_m^{\otimes 2}(\vb)}^2/ \rank \!\Pi_m^{\otimes 2}$
and let $\oP(\vb)$ denote the distribution $\oP(\vb) = \Exp_m[P_m(\vb)]$.
We separate $\norm{P_m - \oP}_1$ into contributions from basis vectors outside and inside
$B_L \cup \bad$:
\begin{equation}
\label{eq:sep2}
\norm{P_m - \oP}_1 
= \sum_{\vb \notin B_L \cup \bad} \abs{P_m(\vb) - \oP(\vb)}
+ \sum_{\vb \in B_L \cup \bad} \abs{P_m(\vb) - \oP(\vb)}
\enspace .
\end{equation}
The first sum is taken only over vectors $\vb$ for which
$$
\abs{ \norm{ \Pi_m^{\otimes2}(\vb) }^2 - \Exp_m \norm{ \Pi_m^{\otimes2}(\vb) }^2 } 
< e^{-(\alpha/3) \sqrt{n}/\ln n} \enspace . 
$$
Then conditioning on $E_0$ and $E_1$, 
the rank estimate of~\eqref{eq:approx-rank2} implies that
\begin{equation}
\label{eq:out2}
\sum_{\vb \notin B_L \cup \bad}
\abs{P_m(\vb) - \oP(\vb)} 
\leq \frac{e^{-(\alpha/3) \sqrt{n}/\ln n}}{\rank \Pi_m^{\otimes 2}} \cdot |B| 
= \frac{4 e^{-(\alpha/3) \sqrt{n}/\ln n}}{1 + e^{-\Omega(n)}} 
< 8 e^{-(\alpha/3) \sqrt{n}/\ln n}
\enspace . 
\end{equation}

On the other hand, conditioning on $E_0$ and $E_1$ we have 
\[ \abs{B_L \cup \bad} \leq \left( 9 e^{-(\beta/2) \sqrt{n}} + e^{-(\alpha/6) \sqrt{n}/\ln n} \right) |B| 
< 2 e^{-(\alpha/6) \sqrt{n}/\ln n} |B| 
\enspace , 
\]
and then~\eqref{eq:approx-rank2} and~\eqref{eq:exp-2} imply that the total expected probability of the basis vectors in $B_L \cup \bad$ is
\begin{align}
  \sum_{\vb \in B_L \cup \bad} \oP(\vb) 
  &= \sum_{\vb \in B_L \cup \bad} \frac{\Exp_m \norm{\Pi_m^{\otimes2}(\vb)}^2}{\rank \Pi_m^{\otimes 2}} 
  \leq \frac{ \abs{B_L \cup \bad} }{\rank \Pi_m^{\otimes 2}}\cdot \left(\frac{1}{4} + 3e^{-\alpha\sqrt{n}/ \ln
      n}\right)
      \nonumber \\
  &\leq 2e^{-(\alpha/6) \sqrt{n}/\ln n} (1 + o(1)) 
  < 3e^{-(\alpha/6) \sqrt{n}/\ln n}
  \enspace . 
  \label{eq:nottooshabby}
\end{align}
Then we must have
$$
\sum_{\vb \notin B_L \cup \bad} \oP(\vb) > 1 - 3e^{-(\alpha/6) \sqrt{n}/\ln n}
$$
and hence, by~\eqref{eq:out2},
$$
\sum_{\vb \notin B_L \cup \bad} P_m(\vb) 
> 1 - 3e^{-(\alpha/6) \sqrt{n}/\ln n} - 8 e^{-(\alpha/3) \sqrt{n}/\ln n}
> 1 - 4e^{-(\alpha/6) \sqrt{n}/\ln n}
$$
and so
$$
\sum_{\vb \in B_L \cup \bad} P_m(\vb) < 4e^{-(\alpha/6) \sqrt{n}/\ln n}
\enspace .
$$
Combining this with~\eqref{eq:nottooshabby} bounds the second sum in~\eqref{eq:sep2}, 
\begin{equation}
\label{eq:in2}
\sum_{\vb \in B_L \cup \bad} \abs{P_m(\vb) - \oP(\vb)} 
< 7e^{-(\alpha/6) \sqrt{n}/\ln n}  \enspace . 
\end{equation}
Then combining~\eqref{eq:sep2}, \eqref{eq:out2} and~\eqref{eq:in2}, 
$$
\norm{P_m - \oP}_1 < 8 e^{-(\alpha/6) \sqrt{n}/\ln n} 
$$
with probability at least $\Pr[E_0 \land E_1]  \geq 1-e^{-\Omega(\sqrt{n})}-e^{-(\alpha/6)\sqrt{n}/\ln n} > 1 - 2 e^{-(\alpha/6)\sqrt{n}/\ln n}$.

Finally, it remains to be proved that $\oP$ is, with high probability, close to the uniform distribution $U$ on $B$.  But this follows from~\eqref{eq:approx-rank2} and~\eqref{eq:exp-2}; conditioning on $E_0$, we have
\[ 
\norm{\oP-U}_1 \leq \sum_{\vb \in B} 
\abs{\frac{\Exp_m \norm{\Pi_m^{\otimes 2} \vb}^2}{\rank \Pi_m^{\otimes 2}}-\frac{1}{|B|}}
< 12 e^{-\alpha \sqrt{n}/\ln n} (1+e^{-\Omega(n)}) \enspace . 
\]
We complete the proof by setting $\delta < \alpha/6$ and invoking the triangle inequality.
\end{proof}

\section{Conclusion}

The reader will notice that our current machinery cannot extend to
three or more registers when applied to the symmetric group, as the
representations of $S_n$ have typical dimension equal to $(n!)^{1/2 -
  o(1)}$.  However, we have been very pessimistic in our analysis; in
particular, we have assumed that vectors of the form $\vb \otimes \vb$
project into low-dimensional representations, $S^\nu$ with $\nu \in \Lambda$,
as much as possible. Perhaps a more detailed understanding of how
these vectors lie inside the decomposition of $V(\vrho,\vsigma)$ into
irreducibles would allow one to prove that this hidden subgroup
problem requires entangled measurements over $\Omega(\log |G|)=\Omega(n \log
n)$ coset states. Therefore, we make the following conjecture.

\begin{conjecture}
Let $B=\{\vb\}$ with weights $\{ a_\vb \}$ be a complete frame for 
$S^{\lambda_1} \otimes \cdots \otimes S^{\lambda_k}$.  
Given the hidden subgroup $H=\{1,m\}$ where $m$ is chosen uniformly at random 
from $M$, and a coset state $\ket{c_1 H} \otimes \cdots \otimes \ket{c_k H}$
on $k$ registers, let $P_m(\vb)$ be the
probability that we observe the vector $\vb$ conditioned on having observed 
the representation names $\{ S^{\lambda_i} \}$, and let $U$ be the natural distribution 
on $B$.  Then for all $c > 0$, with probability $1-o(n^{-c})$ in $m$ and $\{ S^{\lambda_i} \}$, 
we have
\[ \norm{ P_m - U }_1 = o(n^{-c}) \]
unless $k=\Omega(n \log n)$.
\end{conjecture}

\section*{Acknowledgments.}  
This work was supported by NSF grants CCR-0093065, PHY-0200909,
EIA-0218443, EIA-0218563, CCR-0220070, and CCR-0220264.  
We are grateful to Denis Th{\'e}rien, McGill University, and Bellairs Research Institute 
for organizing a workshop at which this work began; 
to Dorit Aharonov, Daniel Rockmore, Leonard Schulman, and Umesh Vazirani for helpful conversations; 
to Sean Hallgren, Martin R\"otteler and Pranab Sen for identifying a flaw in an earlier draft; 
and to Tracy Conrad and Sally Milius for their support and tolerance.  
C.M.\ also thanks Rosemary Moore for providing a 
larger perspective.

\end{document}